\documentclass{egpubl}
\usepackage{EGUKCGVC2021}
\usepackage{xcolor}

\WsPaper           

\usepackage[T1]{fontenc}
\usepackage{dfadobe}  

\usepackage{cite}  
\BibtexOrBiblatex
\PrintedOrElectronic
\ifpdf \usepackage[pdftex]{graphicx} \pdfcompresslevel=9
\else \usepackage[dvips]{graphicx} \fi

\usepackage{egweblnk} 

\graphicspath{{pictures/}}

\usepackage{todonotes}
\usepackage{xfrac}
\usepackage{enumitem}
\usepackage{colortbl} 

\title[3D Visualisations Should Not be Displayed Alone -- Encouraging a Need for Multivocality in Visualisation]%
      {3D Visualisations Should Not be Displayed Alone -- Encouraging a Need for Multivocality in Visualisation
      }
\author[J.\,C. Roberts  \& ]
{\parbox{\textwidth}{\centering J.\,C. Roberts\thanks{Email j.c.roberts@bangor.ac.uk}$^{1}$\orcid{0000-0001-7718-3181},
J.\,W. Mearman\thanks{Email j.w.mearman@bangor.ac.uk}$^{1}$\orcid{0000-0002-9315-9760}, 
P.\,W.\,S. Butcher\thanks{Email p.butcher@bangor.ac.uk}$^{1}$\orcid{0000-0002-3361-627X},
H. M. Al-Maneea\thanks{Email h.m.almaneea@bangor.ac.uk}$^{1,2}$\orcid{0000-0002-9841-1486}
and P.\,D. Ritsos\thanks{Email p.ritsos@bangor.ac.uk}$^{1}$\orcid{0000-0001-9308-3885}
        }
        \\
{\parbox{\textwidth}{\centering $^1$Bangor University, UK\\
          $^2$University of Basrah, Iraq\vspace{10mm}
       }
}
}

\begin{document}


\maketitle
\begin{abstract}
We believe that 3D visualisations should not be used alone; by coincidentally displaying alternative views the user can gain the best understanding of all situations. The different presentations signify manifold meanings and afford different tasks. Natural 3D worlds implicitly tell many stories. For instance, walking into a living room, seeing the TV, types of magazines, pictures on the wall, tells us much about the occupiers: their occupation, standards of living, taste in design, whether they have kids, and so on. How can we similarly create rich and diverse 3D visualisation presentations? How can we create visualisations that allow people to understand different stories from the data? In a multivariate 2D visualisation a developer may coordinate and link many views together to provide exploratory visualisation functionality. But how can this be achieved in 3D and in immersive visualisations? Different visualisation types, each have specific uses, and each has the potential to tell or evoke a different story. Through several use-cases, we discuss challenges of 3D visualisation, and present our argument for concurrent and coordinated visualisations of alternative styles, and encourage developers to consider using alternative representations with any 3D view, even if that view is displayed in a virtual, augmented or mixed reality setup.


\end{abstract}  
\section{Introduction} 
In this paper we lay the foundations of our hypothesis: that when a developer is displaying data in 3D they should also use other depiction methods alongside. They need to use different strategies that accompany each other to enable people to understand the richness of the data, see it from different viewpoints, and deeply understand complexities within it. A single data-visualisation can be used to tell different stories. People can observe, maximum or minimum values, averages, compare data points to known values, and so on, from one visualisation depiction. But when several visualisation depictions are used together, people can view the data from different perspectives. Alternative presentations allow people to understand different points of view, see the data in different ways, or fill gaps of knowledge or biases that one view may give. 

In many cases, it may be possible to coordinate the user-manipulation of each of the views~\cite{Roberts2007}. Through methods such as linked brushing or linked navigation the user can then understand how the information in one view is displayed in another view. But sometimes it is not obvious how to create multiview solutions, or how to link the information from one view to another. For instance, tangible visualisations (printed on a 3D printer) can be used as a user-interface tool, but it may not be clear how to coincidentally display other information, or to `link' the manipulation of these objects directly with information in other views. 

Since the early days of visualisation research, developers have created three-dimensional visualisations. Users perceive 3D through depth perception~\cite{cutting1995perceiving,mehrabi2013making} and understand data through visual cues; visualisation designers map values to attributes of 3D geometry (position, size, shape, colour and so on). Perhaps the data to be examined is multivariate, and maybe one or more of the dimensions are spatial, or it is possible that the developer wants to create an immersive data presentation. Whatever the reason, three-dimensional visualisations can enable users to become immersed in data. 3D Visualisations range from medical reconstructions, depictions of fluid flowing over wings, to three-dimensional displays of network diagrams, charts and plots. They can be displayed on a traditional two-dimensional monitor (using computer graphics rendering techniques), augmented onto live video, or stereo hardware to allow users to perceive depth. Data that has a natural spatial dimension may be best presented as a 3D depiction, while other data is more abstract and it is better displayed in a series of 2D plots and charts. But for some datasets, and some applications, it is not always clear if a developer should depict the data using 2D or 3D views. 

Recently, especially due to the price drop of head-mounted displays (HMD), many researchers have explored how to visualise data in these immersive worlds~\cite{marriott2018immersive}. Consequently, it is timely to critically think about the design and use of three-dimensional visualisations, and the challenges that surround them.  We use a case-study approach, and explain several examples where we have developed data-visualisation tools that incorporate 3D visualisations alongside 2D views and other representation styles. We use these visualisations to present alternative ideas, and allow users to investigate and observe multiple stories from the data. Following the case-studies we discuss the future opportunities for research.

\section{Related Work}
The third dimension has been used by many visualisation developers to display data. Since the early 1990s researchers have used the third dimension to ``shift some of the user’s cognitive load to the human perceptual system''~\cite{RobertsonETAL1991}. Understanding 3D worlds relies on humans to perceive depth~\cite{cutting1995perceiving}. Depth perception can be modelled using monocular cues or displayed in a stereo device~\cite{Butcher-et-al-TVCG-2021}.  When using monocular cues the image can be displayed on a 2D monitor, or augmented onto a video stream. This is why developers sometimes call these images 2\sfrac{1}{2}D~\cite{DixonFitzhughAleva2009}. Users understand that it is a 3D model because of different visual cues, such as occlusion, rotation, shadows, shading and so on. Stereo devices use two difference images that are displayed separately to each eye (e.g., head-mounted display, stereo glasses, or auto-stereoscopic display device).
In addition, there is a third option with data visualisation, where different dimensions, different aspects of the data, or pairs of dimensions, can be displayed in separate juxtaposed views~\cite{Roberts2007,RobertsETALMultiViewMeanings2019}. For instance, these could be side-by-side views, dual views, or three view systems~\cite{RobertsBracketing2004}. There are different view types, that could be used together to help users understand the data. Different visualisations could be lists, table views, matrix plots, SPLOMs, parallel coordinate plots or the dimension reduced using a mathematical dimension reduction algorithm (e.g., principal component analysis, PCA). 

The first challenge, when faced with a new dataset, is to understand the makeup of the data and ascertain appropriate visual mappings. Mappings that exchange data values into appropriate visual artefacts that can be perceived.  The second challenge is to understand how the information will be displayed and what technology will be used to display it.

\subsection{Mapping}
Mapping data to the visual display is obviously a key aspect to the visualisation design, but to create appropriate mappings the developer needs to understand the data they wish to visualise. Shneiderman~\cite{Shneiderman1996} describes the common data types of 1- 2- 3-dimensional data, temporal and multidimensional data, and tree and network data. There is an explicit difference between the type of visualisations that can be made from each of the types of data. For instance, volumetric data (such as from a medical scan) can be naturally displayed in three-dimensions, and it is clear to see the utility of placing the data into a volumentric visualisation style. Multidimensional data, that does not have any spatial coordinates, could be projected into a three-dimensional space as a three-dimensional scatterplot, or displayed in a scatterplot-matrix view in two-dimensions. Or, positional data, from geopositional data (such as buildings on a map) could also be projected into three-dimensional space, or located on a two-dimensional map.
It is clear that there are benefits to displaying objects in three-dimensions. Especially if the data is representing something that is three-dimensional in the real-world. Shneiderman~\cite{Shneiderman2003} writes ``for some computer-based tasks, pure 3D representations are clearly helpful and have become major industries: medical imagery, architectural drawings''.

There are some areas of interactive entertainment that successfully employ 3D. For instance, games developers have created many popular 3D games, but rather than totally mimicking reality they have compromised, and adapted the fidelity of the world representation~\cite{Williams-et-al-MDPI-2020}. Many 3D games employ a third-person view, with the user being able to see an avatar representation of themselves. Obviously the interaction is different to reality, but the adaption allows the user to view themselves in the game and control the character more easily. There are always different influences that govern and shape the creation of different visualisation designs: the data certainly governs what is possible, but the user's experience and their own knowledge effects the end design, and also the application area and any traditions or standards that a domain may expect or impose~\cite{RobertsETALBOOKFDS}.

Sometimes the visualisation designer may add, or present  data using three-dimensional cues, where the data does not include any spatial value. For instance, it is common to receive an end-of-year report from a company with statistical information displayed in 3D bar charts or 3D pie charts. In this case the third-dimension is used for effect and does not depict any data. While these may look beautiful, the third dimension does not add any value to this information. This third dimension is useless -- in terms of giving the user an understanding of the data. This becomes \textit{chartjunk}~\cite{tufte1983visual}, and is often judged to be bad-practice. However, recent work has started to discover that in some situations, there is worth to using chartjunk. For example, Borga et al.\ \cite{Borgo_etal2012} explain that embellishments helped users to perform better at memory tasks. Not only have researchers looked at the use of 3D chartjunk, but also to the effectiveness of 3D visualisations themselves. 

There are situations where three-dimensional presentations are not suitable due to the task that is required to perform~\cite{Shneiderman1996}. Placing a list of objects (such as file names) on a virtual 3D bookcase, may seem attractive and beautiful to the designer, but actually a list of alphabetically ordered names that a user can re-order in their own way, would enable the user to better search the data.  Consequently, there are many examples of datasets that could be displayed in 3D but would be better to visualise in a 2D plot. For instance, data of two variables, with a category and a value, can be displayed in a bar chart. Data with dates can be displayed on a timeline. Relational data, such as person-to-person transmission in a pandemic, could be displayed in a tree or network visualisation and could be displayed in 3D, but may be better in a 2D projection. In fact, each of these different visual depictions have specific uses and afford specific types of interaction. For example, 2D views are useful to allow the user to select items, whereas 3D views can allow people to perceive information in a location. The purpose of the visualisation can influence if 3D is suitable. The purpose could be to \textit{explore}, \textit{explain} or \textit{present} data~\cite{RobertsETAL2018EVF,RobertsETAL2021ARTEMUS}. For instance, one of the views in a coordinated and multiple view setup could be 3D. On other occasions it could be clearer to explain a process in 2D, whereas in another situation a photograph of the 3D object may allow it to be quickly recognised.

Another challenge with 3D is that objects can become occluded. Parts of the visualisation could be contained within other objects or obscured from the observer from a particular viewpoint, or objects could be mapped to the same spatial location. To help overcome these challenges developers have created several different solution. For example, animation and movement are often used to help users understand 3D datasets. By moving the objects or rotating the view, not only does the viewer understand that it is a 3D object, but problems from viewpoint occlusion can be mitigated. Focus and context or distortion techniques~\cite{LeungApperley1994} such as used with perspective wall~\cite{MitchellKennedy1996} or object separation~\cite{roberts2002regular},
or worlds within worlds~\cite{FeinerBeshers1990} can all help overcome occlusion issues and display many objects in the scene. Finally, 3D can help to overcome field of view issues, which could be useful in immersive contexts.  For example, Robertson et al.\ \cite{RobertsonETAL1991} present advantages of 3D in  the context of a small screen real-estate.

It is clear that there are some situations where 3D can help, while in other situations a 2D view would be better. Work by Cockburn and McKenzie~\cite{CockburnMcKenzie2001,Cockburn2002}, focusing on a memory task, compared 2D and 3D designs. Users searched for document icons that were arranged in 2D, 2\sfrac{1}{2}D or 3D designs. They found that users were slower in the 3D interfaces than the 2D, and that virtual interfaces provided the slowest times. This certainly fuels the negativity surrounding the use of 3D. However, later on Cockburn and McKenzie~\cite{cockburn2004revisiting} followup their earlier work, by focusing on spatial memory, saying that perspective did not make any difference to how well participants recalled the location of letters or flags. Interestingly, they conclude by saying ``it remains unclear whether a perfect computer-based implementation of 3D would produce spatial memory advantages or disadvantages for 3D''. Their research also showed that users seem to prefer the more physical interfaces. 

\subsection{Display and interaction technologies}
Traditionally many interface engineers adopt metaphors to help users navigate the information. Metaphors have long-been used by designers to help users empathise and more easily understand user-interfaces~\cite{roberts2014novel}. By using a metaphor that is well known to users, they will be able to implicitly understand how to manipulate and understand the visual interface and thus the presented data. Early work on user-interface design clearly was inspired by the world around us. For instance, everyday we use the pervasive desktop metaphor, and drag-and-drop files into a virtual trash-can to delete them, or move files into a virtual folder to archive them. Many of these metaphor-based designs are naturally 3D. This approach often creates visualisation designs that are beautiful. Often this ideology works well with high-dimensional data~\cite{marriott2018immersive}. However, it is not only the natural world that can be inspiration for these different designs; designs can be non-physical, visualisation inspired, man-made or natural (nature inspired)~\cite{roberts2014novel}. While many of these designs are implicitly 3D, because they are taken from the natural world (such as ConeTree~\cite{Robertsonetal1991ConeTrees} or hierarchy based visualisation of software~\cite{BalzerDeussen2004})  it is clear that the designers do not restrict themselves to keeping a 3D implementation, and inspiration from (say) nature can also be projected into 2D~\cite{mccormack2018multisensory}. 

One of the challenges against using 3D visualisations is they are still dominated by interfaces that are 2D in nature. Mice, touch screens or pen-based interfaces that have influenced the visualisation field, and these interaction styles are all predominantly 2D. Virtual reality publications have been considering 3D for some time, for instance Dachselt and H\"{u}bner~\cite{DachseltHubner2007} survey 3D menus.  Teyseyre and Campo~\cite{TeyseyreCampo2009} in their review of 3D interfaces for software visualisation  write ``once we turn them into post-WIMP interfaces and adopt specialized hardware \ldots 3D techniques may have a substantial effect on software visualisation''. Endeavouring to create novel designs is difficult. Inspiration for designs can thus come from different aspects of our lives~\cite{Roberts_etal2014Harnessing}. We live in a 3D world, and therefore we would assume that many of the interfaces and visualisations that we create would be naturally three-dimensional. Maybe because many of our input interface technologies are predominantly 2D (mouse positions, touch screens) and much of our output technologies are also 2D (such as LCD/LED screens, data projectors etc.) we have not seen too many true 3D visualisation capabilities; most immersive (stereo) visualisations still use bar charts, scatterplots, graphs and plots and so on. But does stereo help? Ware and Mitchell~\cite{Ware2005ReevaluatingSA} demonstrated, when evaluating stereo, kinetic depth and using 3D tubes instead of lines to display links in a 3D graph visualisation, depiction of graphs, that there was a greater benefit for 3D viewing. 

Several recent technologies are transformational for visualisation research. These technologies allow developers to move away from relying on WIMP interfaces and explore new styles of interaction~\cite{RobertsWalker2010}. These interfaces move `beyond the desktop'~\cite{Fuchs1998,JensenDragicevic2013,Bongshin_etal2012,Roberts_etal2014b} even becoming more natural and \textit{fluid})~\cite{Elmqvist_etal2011}. For example, 3D printing technologies have become extremely cheap (Makerbot or Velleman printers are now affordable by hobbyists) which can be used to easily make tangible (3D printed) objects~\cite{Spindler_etal2010}. These tangible objects become \textit{props}~\cite{KruszynskiLiere2009} as different input devices, or become conversational pieces around which a discussion with a group of people can take place (as per the 3D printed objects in our heritage case-study, in Section~\ref{sec:heritage}). Haptic devices (such as the Phantom or Omni~\cite{PaneelsRoberts2010,Paneels_etal2013}) enable visualisations now to be dynamically felt. There is a clear move to integrate more senses other than sight~\cite{RobertsWalker2010}, sound and touch~\cite{Paneels_etal2013}, and modalities such as smell~\cite{BatchETAL_smell2020} are becoming possible. These will certainly continue to develop and designers will invent many more novel interaction devices. In fact, in our work, we have been using tangible devices to display and manipulate the data. 3D printed objects become tangible interaction devices, and act as data surrogates for the real object. 
However, while on the one hand there is a move away from the desktop, it is also clear to see that most visualisations use several methods together. For instance, a scatter plot shows the data positioned on xy coordinates, has an axis to give the information context, adds text labels to name each object (otherwise the user would not understand what the visualisation is saying). Likewise, we postulate that, even when we are displaying the data using 3D that developers need to add appropriate context information. These could be axis, legends, associated scales, and other reference information to allow people to fully understand the information that is being displayed. 

\begin{figure}[b]
	\centering
    \includegraphics[width=1\linewidth]{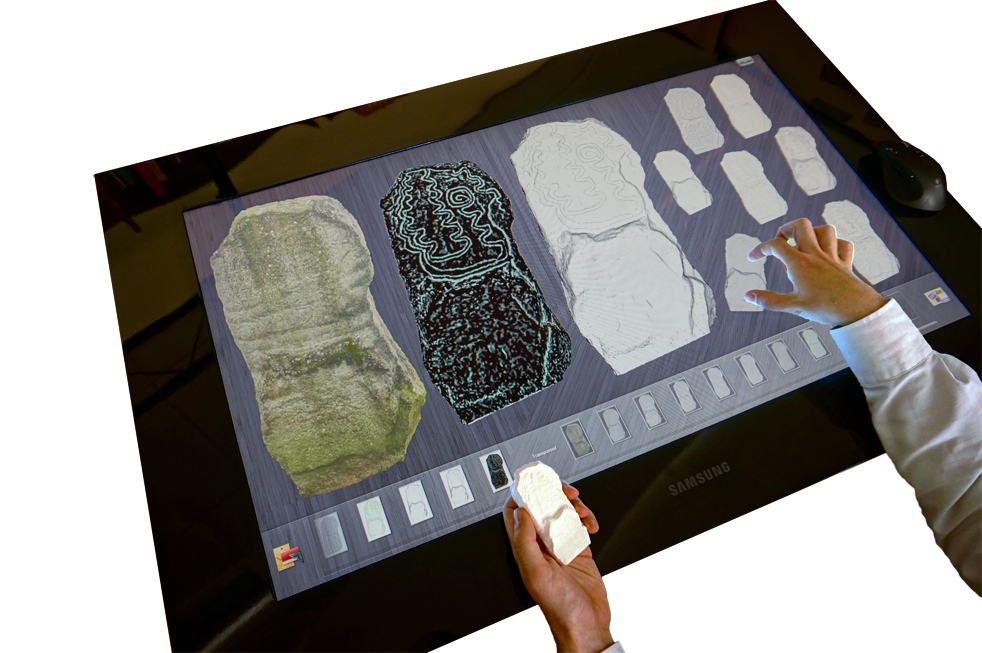}
    \caption{Images of the prehistoric standing stone, at Bryn Celli Ddu North Wales site, displayed on the touch table. Showing three large 3D pictures of the standing stone (fully textured and rendered, line rendered version to enhance the rock carvings, and the plain shaded version), along with smaller alternative depictions. The user is holding the tangible representation of the standing stone.\label{bcd}}%
   \vspace*{-1.2em}
\end{figure}

\section{Case study -- Cultural Heritage Data}
\label{sec:heritage}
There are many researchers who wish to gather digital representations of tangible heritage assets. One of the reasons is that many of these heritage sites are deteriorating. Wind, snow, rain and even human intervention, can all effect these old sites. Therefore conservationists wish to survey and scan these sites such to create digital representations. Furthermore, these digital assets can then be analysed and investigated further; they can be better compared.

In \href{http://heritageTogether.org}{heritageTogether.org}, using a citizen science approach, members of the public photograph standing stones, dolmen, burial cairns and so on, which are then changed to 3D models through a photogrammetry server~\cite{MilesETAL2015,GriffithsETAL2015,GriffithsETAL2015NUMBER2}. These are naturally three-dimensional models. However, we also store (and therefore can reference) statistical information, historical records of excavation, location data and maps, archival photographs.
The challenge for the archaeologist is that not one three-dimensional model tells the full story. A full-rendered picture of the site, certainly gives the user the perception of scale; but it is difficult to observe detail. It is also difficult to understand quantitative data of soil ph levels or carbon-dating from samples taken from the site when viewing a single rendered view of the site.  What is required is a multiple-view approach~\cite{MilesETAL2016,miles2015alternative}. 

\begin{figure}[t]
	\centering
    \includegraphics[height=4.24cm]{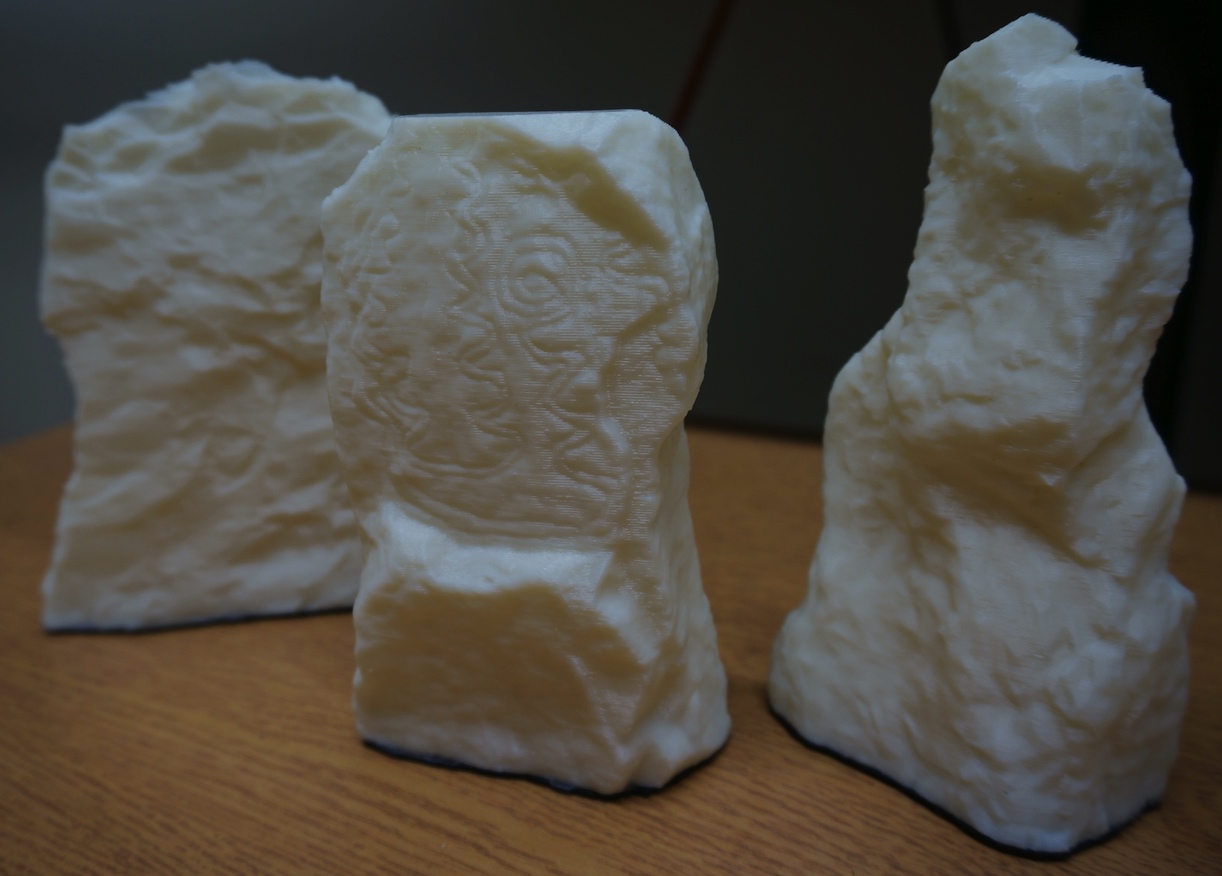}
    \includegraphics[height=4.24cm]{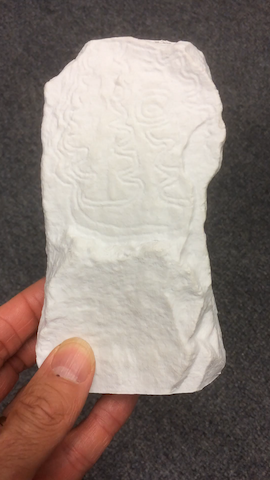} 
    \caption{Several 3D printed models of prehistoric standing stones. Right shows the stone from the Bryn Celli Ddu site in North Wales, that depicts the rock art. \label{fig:printedStones}}\vspace*{-1.2em}
\end{figure}

Our approach is to combine alternative visualisation techniques: graphs and line-plots to demonstrate the statistical data and trends, maps to demonstrate positions and give context and to show the same type of site (prehistoric site) over the landscape; 3D printed models to enable discussion; high-quality rendered images to show detail; and 3D rendered models depicted \textit{in situ} through web-based AR~\cite{Ritsos-et-al-EGGCH2014}. Each of these models enable the user to create a different perception and understanding of the data.
In fact, after sketching different designs~\cite{Roberts_FDS_2016}, we are developing a visualisation tool that integrates renderings, alongside traditional visualisation techniques of line-plots, time-lines, statistical plots etc. to enable the user to associate the spatial data with statistical data and map data. Figure~\ref{bcd} shows our prototype interface with renderings of  
Bryn Celli Ddu. This is a neolithic standing stone which is part of the Atlantic Fringe and contains abstract carvings. Using the SUR40 Samsung table-top display users are able to combine 3D views with 2D statistics, with tangible 3D models (several models are shown in Figure~\ref{fig:printedStones}). Some standing stones have carved patterns. Because of the weathering of the stones and their texturing, the carvings are difficult to observe (either on site, or on the rendered models). However by removing the texture, or rendering the models under different lighting conditions, the carvings become obvious.

\begin{figure}[ht!]
	\begin{center}
	\includegraphics[width=.8\linewidth]{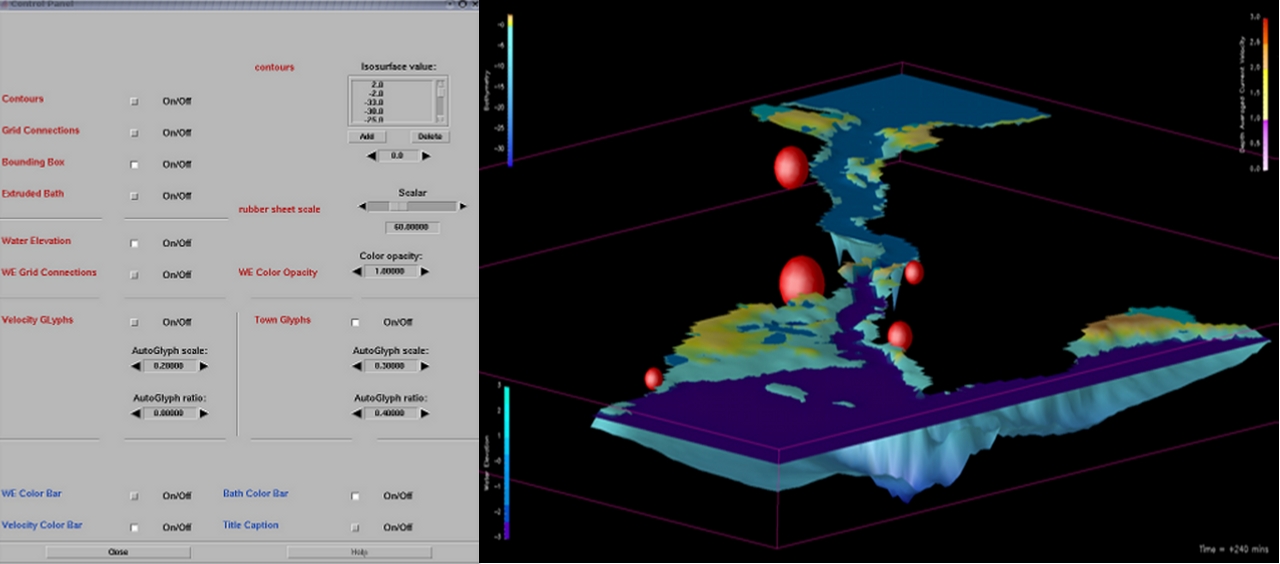}\\[0.5em]%
	\includegraphics[width=.8\linewidth]{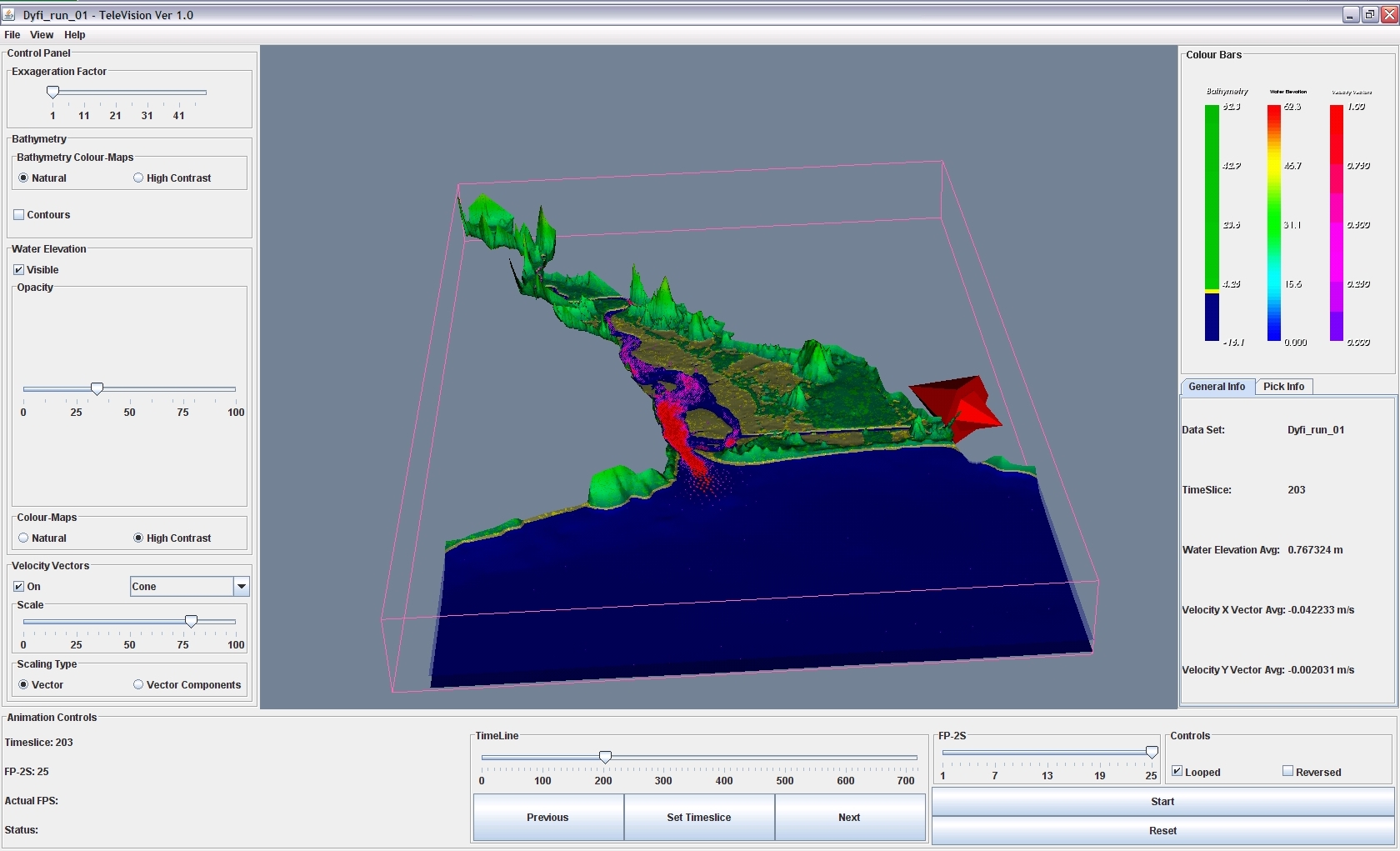}\\[0.5em]%
	\includegraphics[width=.8\linewidth]{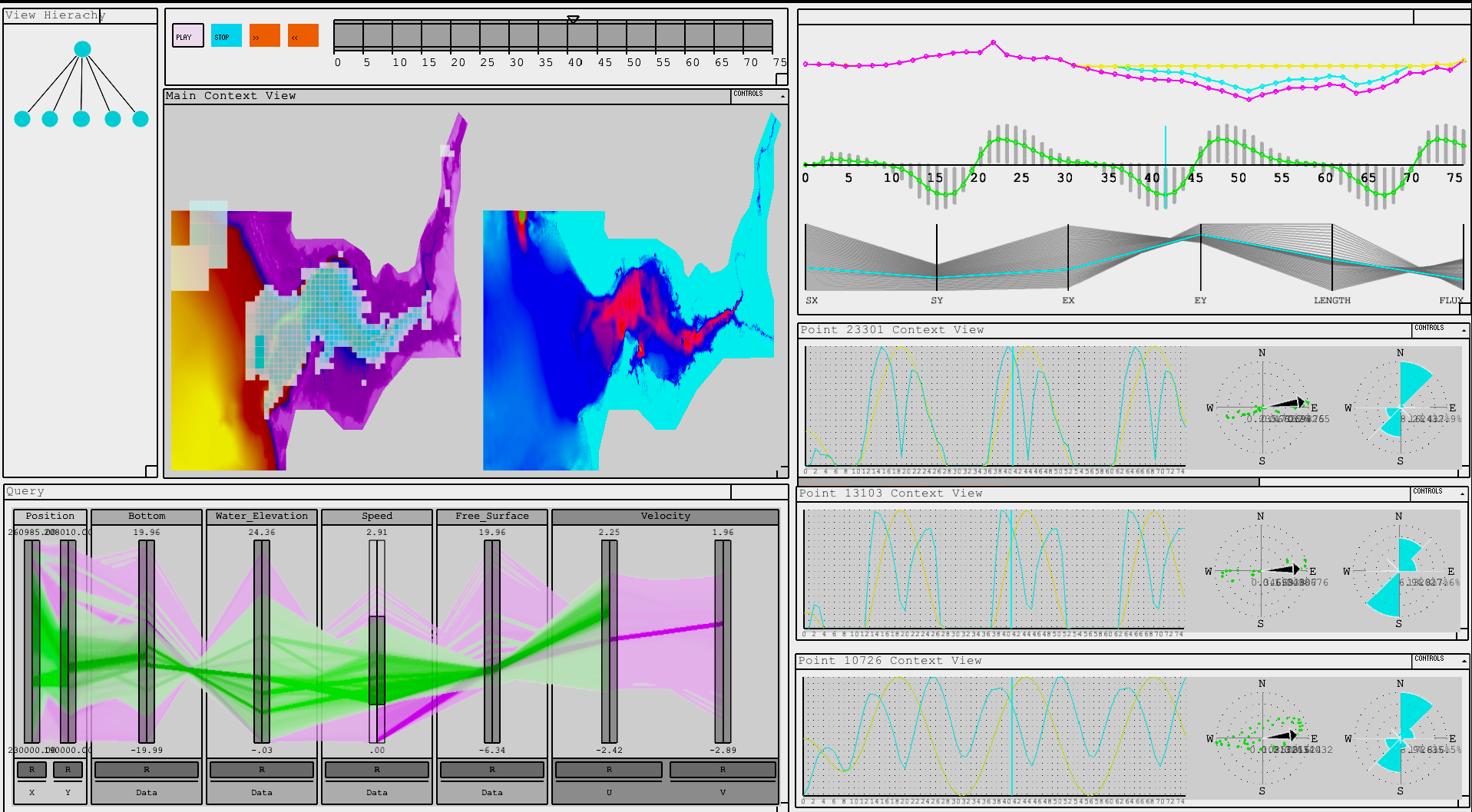}%
    \end{center}
	\caption{We developed several prototypes. The first two (top and center) use VTK and the primary three-dimensional view dominates the interface, with the final version (Vinca) and shown on the bottom, depicts a projection 3D view with many associated coordinated views alongside~\cite{george2014interactive}.\label{Vinca}}
	\vspace*{-1.2em}
\end{figure}

\section{Case study -- oceanographic visualisation}
\label{sec:oceanCaseStudy}
In the second example, we focus on oceanographic data.  Scientists wish to understand how sediment transports up an estuary, understand how sediment affects flooding, and over-topping events, where the sea comes over the sea walls and floods the land, is sometimes due to the movement of silt. 
This data is naturally three-dimensional. It contains positional information and eleven other parameters (including salinity, temperature, velocity). Real-world samples and measurements are taken that feed into the TELEMAC mathematical models.  Our visualisation tool (Vinca~\cite{george2014interactive}), developed in the processing.org library and OpenGL~\cite{GeorgeETAL2010iCove}, provides a coordinated multiple view approach to the visual exploration.

Figure~\ref{Vinca} shows our three prototypes. The top two screenshots show our early prototypes with a single 3D view, with visual information annotated in the 3D space. However through consultation, the oceanographers wanted to be able to take exact measurements, calculate the flux and quantity of water transported by the currents. The final prototype therefore integrated a 3D view coordinated with many other views, including tidal profiles, a parallel coordinate plot of all the data in the system and rose plots. Specific points can be selected and highlighted in x,y,z space, transepts across the estuary can be made in the 3D view to be matched with specific profile plots.

\section{Case study -- Immersive Analytics}
\label{sec:IAcaseStudy}
\begin{figure}[ht!]
	\begin{center}
	\includegraphics[width=.76\linewidth]{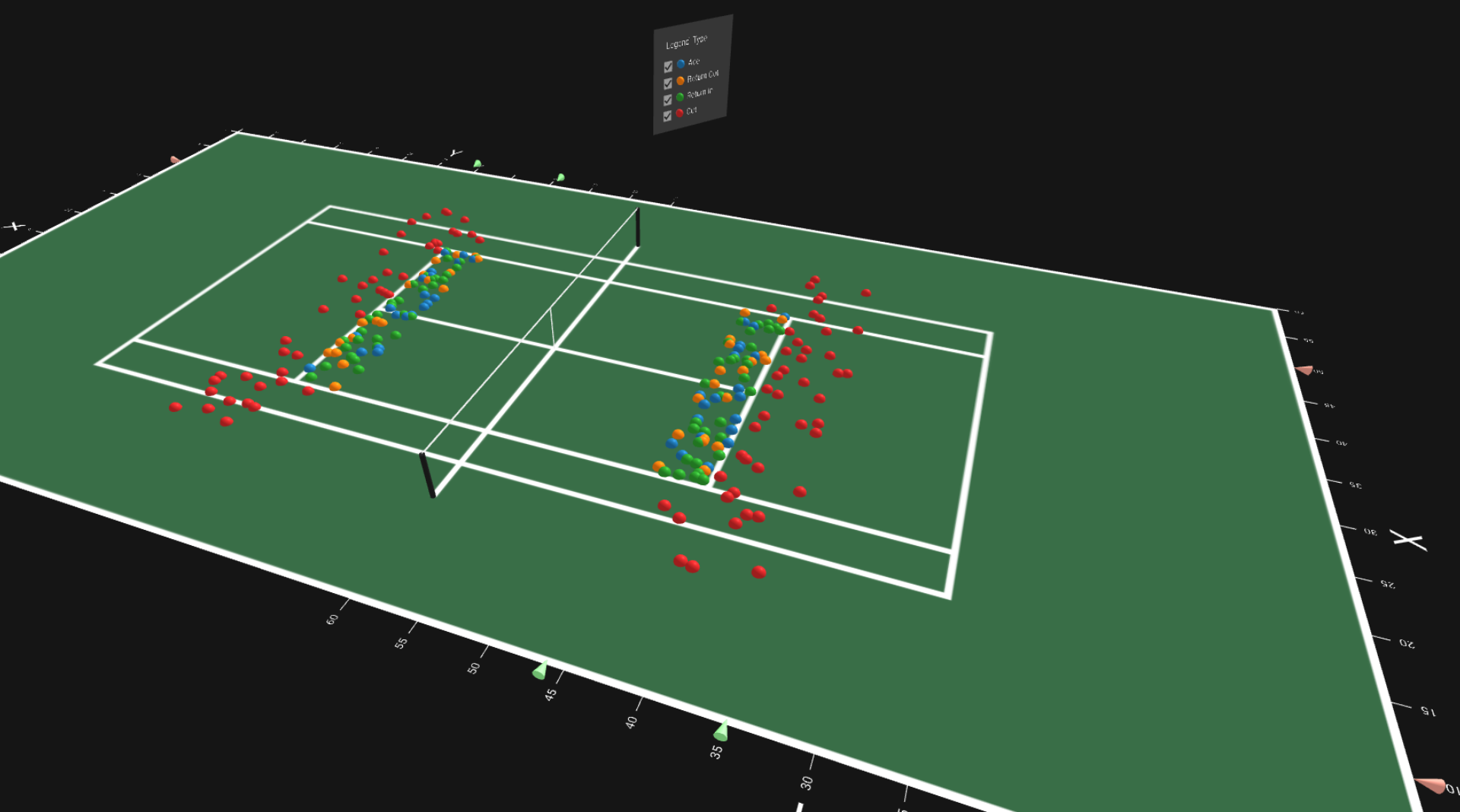}\\[0.2em]%
	\includegraphics[width=.76\linewidth]{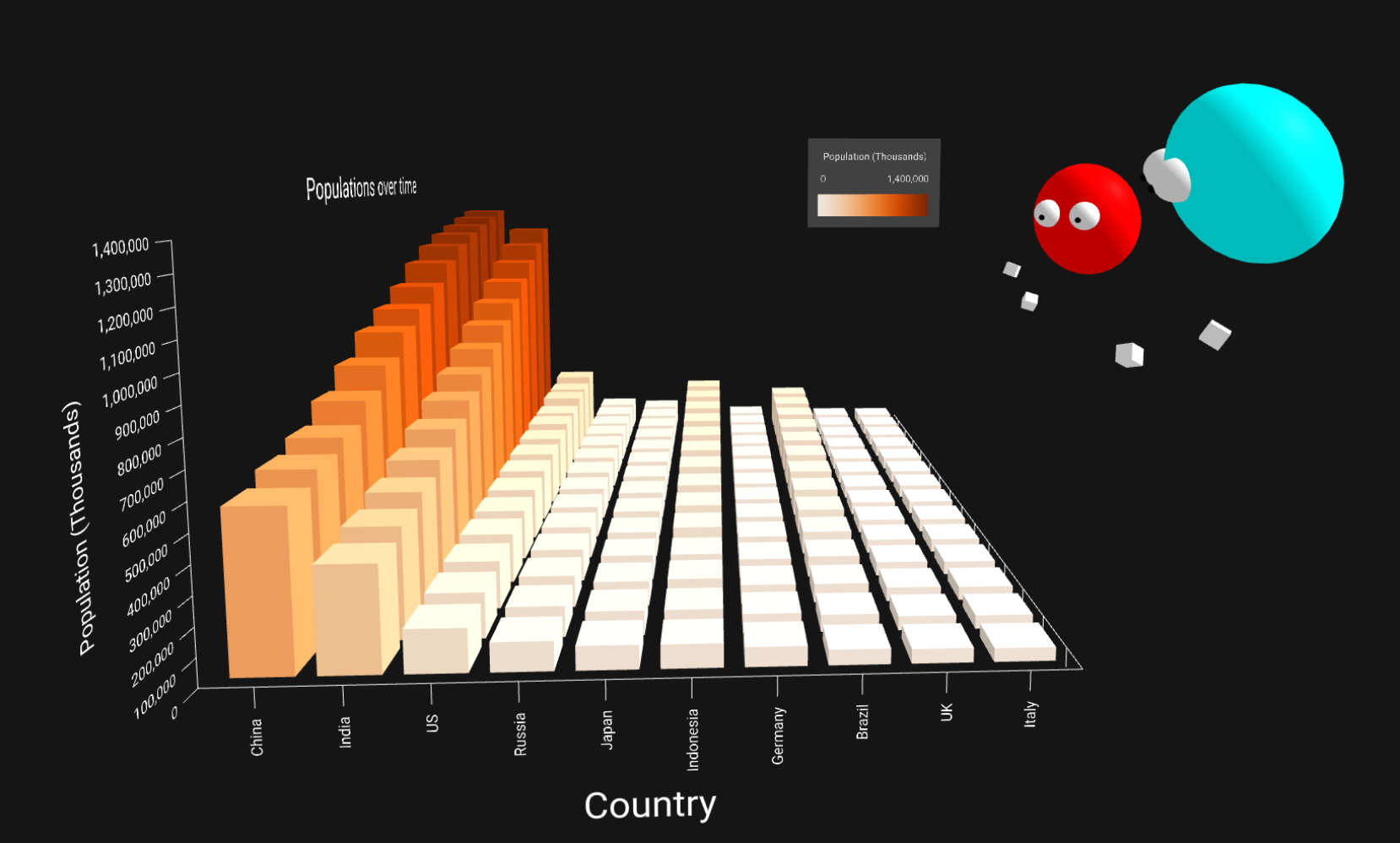}
    \end{center}
	\caption{Example use-cases created with the VRIA framework~\cite{Butcher-et-al-TVCG-2021}. The top image depicts a visualisation of the service game of two tennis players, contextualised with the court props. In the bottom image, in addition to axis and legends, the presence and interactivity of collaborators becomes evident by animating 3D heads, based on viewpoint, and their hands (input from HMD hand controllers).}
	\label{fig:VRIA}
\end{figure}

One of the emerging uses of 3D depictions is in the domain of Immersive Analytics (IA)~\cite{marriott2018immersive}, which builds on the synergy of contemporary XR interfaces, visualization and data science. IA attempts to immerse users in their data by employing novel display and interface technologies for analytical reasoning and decision making, with more advanced flavours introducing multi-sensory~\cite{BatchETAL_smell2020}, and collaborative~\cite{Roberts_etal2014b} set-ups.
In our work on VRIA~\cite{Butcher-et-al-LBW-CHI2019, Butcher-et-al-TVCG-2021}, a Web-based framework that enables the creation of IA experiences using Web technologies, we have observed the importance of 3D depiction for analytical tasks, which are supported by text, axes, filter handlers etc. and from elements that enable contextualisation, such as visual embodiments of data-related objects~\cite{Williams-et-al-VIS-2020}, models and props. These elements not only enhance the user experience of participants in the immersive environment, but more importantly facilitate the analytical process, and often provide a degree of data viscerilisation~\cite{9229242}. For example, when depicting the service game of two tennis players (Figure~\ref{fig:VRIA}, top), the court's outline provides an indication of service patterns, the quality of the game etc.

\begin{figure}[t!]
	\begin{center}
	\includegraphics[width=.49\linewidth]{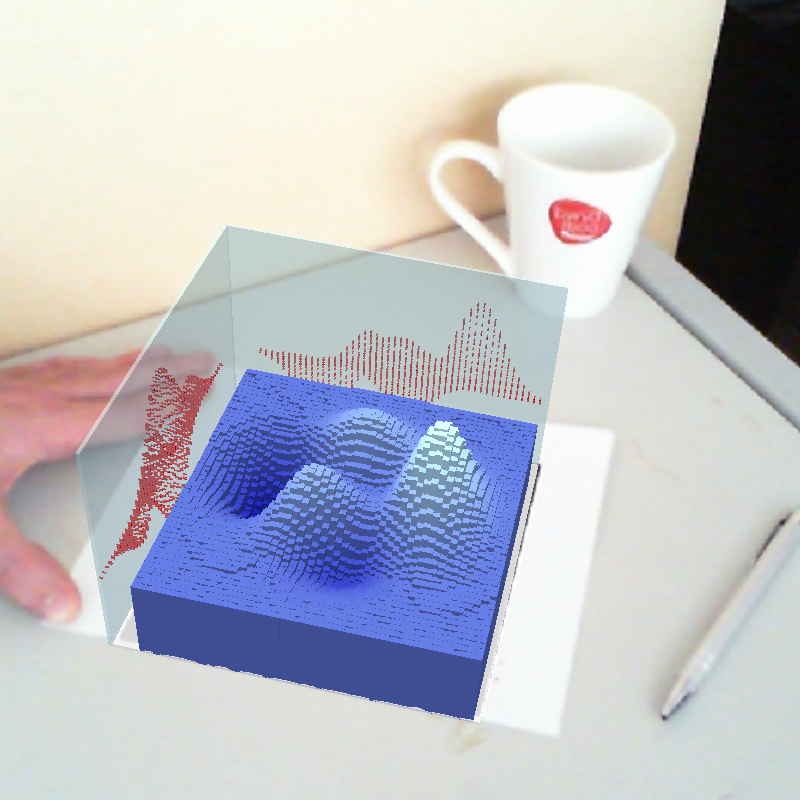}\hfill
	\includegraphics[width=.49\linewidth]{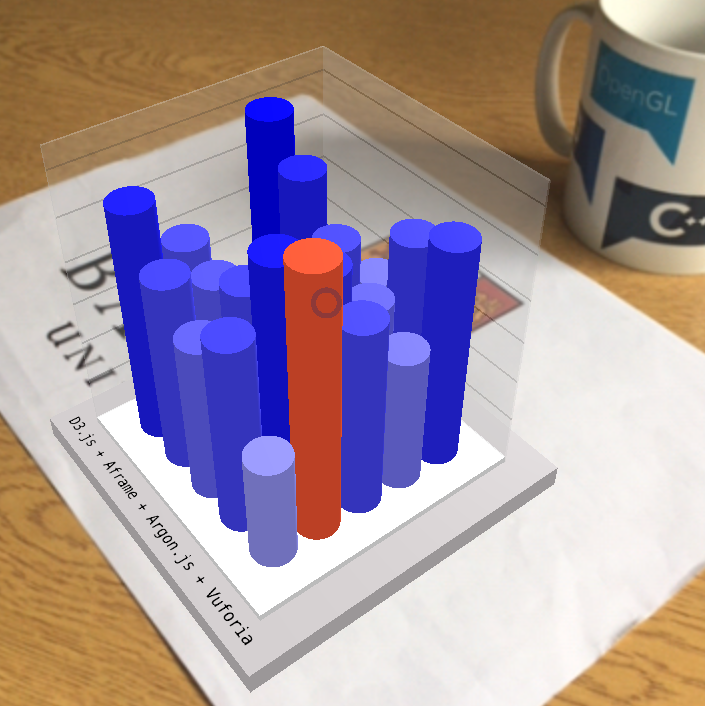}
    \end{center}
	\caption{Situated Analytics prototypes that use Web technologies and can be experienced via standard or mobile browsers~\cite{Ritsos-et-al-IAW-2017, Ritsos-et-al-Poster-VIS2017}. The use of perpendicular semi-transparent guide planes facilitate the understanding of value. However, for both depictions, the absence of textual information hinders the precise understanding of said values. Annotations could be used, but when using a handheld device, targeting may be challenging.}
	\label{fig:XR}
\end{figure}

Another form of contextualisation is in the use of situated analytics, which are essentially immersive analytic systems that use mixed and augmented reality (MR/AR). In this scenario, a 3D depiction can be presented within physical space, or upon a marker object~\cite{Ritsos-et-al-IAW-2017}, that adds context and meaning to the depiction (see Figure~\ref{fig:XR}). In such depictions, the 3D information is evidently not alone, however any additional embellishment or textual information used, must take into account issues such as occlusion of physical objects (especially if these matter for the comprehension of the visualisation), scaling when markers are used for registration, and definition when the background or lighting conditions make the visualization harder to read. The latter of course applies for the main 3D depiction as well.

\definecolor{LightCyan}{rgb}{0.88,1,1}
\definecolor{LightGray}{rgb}{0.8,.8,.8}
\newcolumntype{z}{>{\columncolor{LightCyan}}r}
\newcolumntype{y}{>{\columncolor{LightGray}}r}

\begin{table}[t]
  \centering
  \vspace{0mm}
  \caption{Quantitative data from our multiple view analysis, shows that single-view systems are spread across the years 2012-2018, and form   14\% of all systems. 3-view systems are the most popular, and that 60\% of the systems in our study have 4-views or less.}
  \label{tab:FrequencyOfViews}%
  \resizebox{\columnwidth}{!}{%
    \begin{tabular}{@{}rr@{\hspace{2mm}}r@{\hspace{2mm}}r@{\hspace{2mm}}r@{\hspace{2mm}}r@{\hspace{2mm}}r@{\hspace{2mm}}r@{\hspace{2mm}}rzr}
    \hline
    \textbf{Views} & \textbf{'12}  & \textbf{'13}  & \textbf{'14}  & \textbf{'15}  & \textbf{'16}  & \textbf{'17}  & \textbf{'18}  & \multicolumn{1}{r}{\textbf{Freq.}} & \multicolumn{1}{z}{\textbf{\% Freq.}}&Rank \\ 
    \hline
    \textbf{1} & 10     & 7     & 12    & 12    & 9     & 8     & 11    & 69    & 14.1&3 \\
    \textbf{2 }& 10    & 6     & 7     & 9     & 5     & 11    & 14    & 62    & 12.6&4 \\
    \rowcolor{gray!25}\textbf{3}& 8     & 11    & 6     & 10    & 14    & 16    & 20    & 85    & \textbf{17.3}&1 \\
    \textbf{{4}} & 12    & 8     & 11    & 13    & 10    & 10    & 14    & 78    & 15.9&2 \\
    \textbf{5+} & 23     & 18     & 24     & 29     & 39     & 39    & 35    & 197    & 40.1&5+ \\
    \\ \hline
    \rowcolor{gray!5}\textbf{Total} & \textbf{63}    & \textbf{50}    & \textbf{60}    & \textbf{73}    & \textbf{67}    & \textbf{84}    & \textbf{94}    & \textbf{491}   & \textbf{100.0}\\
    \end{tabular}%
    }
  \vspace{-2mm}
\end{table}%

\section{Case study -- quantification of Multiple Views}
\label{sec:quantitative}
For our final case study, we look to how 3D is used in the broader visualisation literature. As part of a larger quantification project, we analyse visualisation research papers between 2012 and 2018~\cite{RobertsETALMultiViewMeanings2019,al2019towards} published at the IEEE VIS conference (including InfoVis, VAST and SciVis). Through a careful deliberation process, we extract screenshots of visualisation tools and store 491 images of different visualisation tools.  We analyse the layout of the view system, count the views used across years, and evaluate the view type~\cite{al2019towards,ChenETAL_2021}. Table~\ref{tab:FrequencyOfViews} summarises our data. It shows that 60\% of all systems contain 4-views or less, and that the most popular type of systems are 3-view systems. 
We also investigate types of charts used. We treat all views in one list, which following from Natural Language Processing terminology, we call a bag-of-views. We classify them as: bar chart, scatter plot, line chart, heatmap, and so on~\cite{al2019towards}, see Table~\ref{tab:viewTypes}. In our classification, we have two 3D view types: `rendered image' and a general `3D' category. We do have a category labelled `other', which is used to record systems that we cannot classify. Potentially there may be some 3D views within this category, but numbers are low. 

\begin{table}[t]
  \centering
  \vspace{0mm}
  \caption{Quantitative data of the view types. Showing that three-dimensional images are ranked 11th and 20th place (from all images). The table also demonstrates that 3D views are more likely to occur alongside other views. }
  \label{tab:viewTypes}%
  \resizebox{.7\columnwidth}{!}{%
    \begin{tabular}{@{}rr@{\hspace{2mm}}r@{\hspace{2mm}}r@{\hspace{2mm}}r@{\hspace{6mm}}}
    \textbf{Rank}&\textbf{Chart type}&\textbf{1-view}&\textbf{+views}&\textbf{Total}\\\hline
\textbf{1}&bar chart&3&333&\textbf{336}\\
\textbf{2}&scatter plot&7&321&\textbf{328}\\
\textbf{3}&line chart&2&226&\textbf{228}\\
\textbf{4}&heatmap&4&214&\textbf{218}\\
\textbf{5}&node link diagram&12&171&\textbf{183}\\
\textbf{6}&small multiple&0&168&\textbf{168}\\
\textbf{7}&map&8&112&\textbf{120}\\
\textbf{8}&text&0&109&\textbf{109}\\
\textbf{9}&area chart&3&98&\textbf{101}\\
\textbf{10}&other&2&70&\textbf{72}\\
\rowcolor{gray!30}\textbf{11}&rendered image&1&63&\textbf{64}\\
\textbf{12}&parallel coordinate plot&5&57&\textbf{62}\\
\textbf{13}&table&1&57&\textbf{58}\\
\textbf{14}&histogram&0&55&\textbf{55}\\
\textbf{15}&treemap&3&51&\textbf{54}\\
\textbf{16}&pie chart&6&36&\textbf{42}\\
\textbf{17}&hierarchy&2&37&\textbf{39}\\
\textbf{18}&star plot&0&29&\textbf{29}\\
\textbf{19}&timeline&1&27&\textbf{28}\\
\rowcolor{gray!30}\textbf{20}&3D&4&22&\textbf{26}\\
\textbf{21}&matrix&1&23&\textbf{24}\\
\textbf{22}&point chart&1&22&\textbf{23}\\
\textbf{23}&bubble chart&2&20&\textbf{22}\\
\textbf{24}&image&1&17&\textbf{18}\\
\textbf{25}&glyph&0&12&\textbf{12}\\
\textbf{26}&video&0&6&\textbf{6}\\
\hline
\textbf{}&&69&2356&\textbf{2425}\\
       \end{tabular}%
   }
  \vspace{-2mm}
\end{table}%

Our quantitative data analysis provides strong evidence that while 3D is used on its own, it is more often found alongside other views. We find that 3D views are most likely to be shown alongside line charts, text, heatmaps, and scatter plots. Whereas `rendered images' are more likely to be seen with node link diagrams, line charts and start plots. It would seem that developers are trying to overcome some of the challenges of viewing in 3D spaces, such as object occlusion, navigation and searching, by linking the 3D information with other views. Anecdotally the 3D images tend to be used in visualisation papers with a strong scientific content, such as medical visualisation, heritage, and flow visualisation. And in the last five years 3D have been used for immersive and situated analytics. Our analysis does have limitations. We have only classified research papers, and not general visualisations that are found on the Internet (e.g., published on a blog). We may have miss-classified some views, although we went through a rigorous checking process. Some categories may include 3D visualisations. For instance, the category `video', which are often used by the visualisation community, could include 3D visualisations. However any single-view videos would be included in the supplementary material and would not be shown as Figures in the papers. Also some visualisation types could be made into 3D depictions, such as a 3D node link diagram, or 3D pie chart. We labelled 3D node link diagrams in the 3D category, and any 3D pie charts would have been included in the pie chart (as the 3rd dimension is just for visual effects).

\section{Discussion  and conclusions}
Each of our case studies tell a different, but synergistic, story. From the heritage scenario (Section~\ref{sec:heritage}) we learn that each alternative 3D view helps with multivocality. The real standing stones in the field, or virtually on a map, show the lay of the land. The rendered models show the deterioration of the heritage artifacts, which can be stored and compared with captured models of previous years. The physical models become tangible interfaces, and can be passed around a group to engender discussion. 
From the oceanographic case study (Section~\ref{sec:oceanCaseStudy}) we understand that quantitative information is better in 2D, but 3D is required to give context, positional information and allow users to select specific locations. It is easier to select a transept across the estuary in the 3D map view, than on the alternative visualisations. 
From our work in Immersive Analytics (Section~\ref{sec:IAcaseStudy}) we understand the power of visual embodiments, to allow people to innately understand the context of the data. If the 3D view is modelled to look like the real-world (that it represents) then users can quickly understand the context of the information. We also learn that without suitable contextual information (or contextual scales, legends and other metainformation) the data presentation can be meaningless.  
From the quantitative study of 1-view and multiple-view systems used in the literature (Section~\ref{sec:quantitative}) we learn that rendered images and immersive analytics both use 3D views, and that tool developers do put 3D information alongside other information in multiple view systems. 

Subsequently, it is evident from our work and the literature, that 3D is required and used by many visualisation developers. There is a clear need to display information in a spatial way, which in turn allows us to become `immersed' in data. 3D views provide many benefits over 2D. For instance, 
3D views provide location information. Immersed views describe context. 3D models, mimicking reality, enable people to relate quickly to ideas. Tangible views are great to get users discussing about a topic, and can act as a interface device.

However, there is strong and growing evidence that developers need to do more than merely place their visualisation into a three-dimensional picture. Let us imagine looking through an archive and finding an old black-and-white photograph of an early computer gamer. The image tells many stories. The fact that it is black and white tells us that it was taken at a time before modern cameras. The curved cathode-ray-tube screen tells us something about the resolution of computers of the day. The clothes of the operator tells us about their working environment. How can we, as developers create 3D visualisations that contain such detailed information? How can we create visualisations that include subtle cues to tell the story of the data? How can we use shadows, lights, dust, fog, and models themselves that express detailed stories that implicitly express many alternative stories as the black-and-white photo did?

Developers should think long and hard how to overcome some of the challenges of the third-dimension, and how to create information-rich visualisations. These include problems of depth perception in 3D, items being occluded, issues of how to relate information between spatial 3D views another other views (possibly 2D views), and challenges of displaying quantitative values and including relevant scales and legends. For instance, placing a node-link diagram in 3D allows people to view the spatial nature of the information, but without any labels it is not clear what that information displays. A visualisation of bar charts augmented on a video feed, may provide suitable contextual information, but if there are no axis or scales, then values cannot be understood. Indeed, what is clear, is that while 3D is used (as one view) within multiple view systems, it is not clear how to add detailed quantitative information to 3D worlds, when the 3D world is the primary view (e.g., with Immersive Analytics). 

Consequently, there are many open research questions.  What is the best way to overcome occlusion in 3D? Is it best to relate information to 2D views, or add windows in 3D? How should labels be included in 3D views (as a 2d screen projection, or in 3D)? What is the best way to add scales, legends, axis and so on in 3D? What is the best way to integrate tangible objects? 
Many 3D visualisations seem to be extensions of 2D. Perhaps developers are stuck on traditional techniques, 2D scatter plots, 2D display devices, 3D volumes. How can we, as developers, think beyond transferring 2D ideas into 3D, and instead create novel immersive 3D environments, that integrate tangible, natural and fluid interaction? How can we create information-rich visualisations in 3D that tell many stories?

In conclusion, there is much importance to showing 3D, but we believe that 3D visualisations need to be shown with other types of views. That users gain a richer understanding of the information through alternative presentations and multiple views. That visualisation developers should create systems that enable many stories and different viewpoints to naturally be understood from the information presentation. We encourage designers of 3D visualisation systems to think beyond 2D, and rise to the opportunities that 3D displays, immersive environments, and natural interfaces bring to visualisation. 

\section*{Acknowledgements}
We acknowledge the reviewers for their comments and suggestions to improve the paper. We acknowledge the UK Arts and Humanities Research Council (AHRC) for funding \href{http://heritageTogether.org}{heritageTogether.org} (grant \href{https://gtr.ukri.org/projects?ref=AH\%2FL007916\%2F1}{AH/L007916/1}).

\bibliographystyle{eg-alpha-doi} 
\bibliography{vis3dnotalone.bib}       

\end{document}